\documentclass[aps,prb,
onecolumn,
longbibliography,nofootinbib,showpacs,preprintnumbers,superscriptaddress]{revtex4-2}

\usepackage{tikz-cd}

\usepackage{psfrag}
\usepackage{mathrsfs}

\usepackage{amsfonts}
\usepackage{amsmath} 
\usepackage{amsthm}
\usepackage{amssymb} 

\usepackage{xparse} 
\usepackage{empheq}		
\usepackage{cancel}     

\usepackage{bbm} 
\usepackage{commath}

\usepackage{graphicx}
\usepackage{epsfig}
\usepackage{float}
\usepackage{color,array,subfigure}
\usepackage{epstopdf}
\usepackage{esint}

\usepackage{geometry}
\usepackage{mathdesign}

\usepackage[utf8]{inputenc}

\RequirePackage{CJKutf8}

\usepackage[colorlinks=true,  
            citecolor=red,
            urlcolor= blue,                     %
            filecolor=black,
            linktocpage=true,  
            linkcolor=blue,
             ]{hyperref}

\usepackage{dcolumn}
\usepackage{bm}

\newcommand{\beqa}{\begin{eqnarray}}
\newcommand{\eeqa}{\end{eqnarray}}
\newcommand{\bit}{\begin{itemize}}
\newcommand{\eit}{\end{itemize}}

\newcommand{\bsp}{\begin{split}}
\newcommand{\esp}{\end{split}}
\newcommand{\bpm}{\begin{pmatrix}}
\newcommand{\epm}{\end{pmatrix}}
\newcommand{\bbm}{\begin{bmatrix}}
\newcommand{\ebm}{\end{bmatrix}}
\newcommand{\bBm}{\begin{Bmatrix}}
\newcommand{\eBm}{\end{Bmatrix}}
\newcommand{\bvm}{\begin{vmatrix}}
\newcommand{\evm}{\end{vmatrix}}
\newcommand{\bVm}{\begin{Vmatrix}}
\newcommand{\eVm}{\end{Vmatrix}}

\newcommand{\bsel}{{\begin{subequations}\begin{empheq}}}
\newcommand{\bse}{{\begin{subequations}\begin{empheq}[left={\ii}\empheqlbrace]{align}}}
\newcommand{\ese}{{\end{empheq}\end{subequations}}}
\def\bc{\begin{center}}
\def\ec{\end{center}}
\def\bnum{\begin{enumerate} }
\def\enum{\end{enumerate}}

\def\ii{\!\!\!\!\!\!}  
\def\3i{\!\!\!}
\def\2i{\!\!}

\def\ec{{e_c}}

\def\({\left(}
\def\){\right)}

\def\vev#1{\left\langle#1\right\rangle}
\def\>{\rightarrow}

\def\Diracslash#1{\not{\hbox{\kern-4pt $#1$}}}

\def\Dslash{\not{\hbox{\kern-4pt $D$}}}
\def\pslash{\not{\hbox{\kern-4pt $p$}}}
\def\qslash{\not{\hbox{\kern-4pt $q$}}}

\def\lv{\not{\hbox{\kern-4pt $L$}}}
\def\lsim{\mathrel{\raise.3ex\hbox{$<$\kern-.75em\lower1ex\hbox{$\sim$}}}}
\def\gsim{\mathrel{\raise.3ex\hbox{$>$\kern-.75em\lower1ex\hbox{$\sim$}}}}
\def\ifmath#1{\relax\ifmmode #1\else $#1$\fi}

\evensidemargin=\oddsidemargin

\usepackage[normalem]{ulem}  

\renewcommand\sout{\bgroup \color{red} \ULdepth=-.5ex \ULset}

\newcommand\p{\partial} 

\newcommand\rd{{\rm d}}

 \makeatletter

\begin{document}

\begin{CJK}{UTF8}{gbsn}

\title{Effective field theories for gapless phases with fractons via a coset construction}

\author{Yuji~Hirono}
\email{yuji.hirono@gmail.com} 

\affiliation{
Asia Pacific Center for Theoretical Physics, Pohang, Gyeongbuk, 37673, Korea
}
\affiliation{
Department of Physics, Pohang University of Science and Technology, Pohang, Gyeongbuk, 37673, Korea
}
\affiliation{
RIKEN iTHEMS, RIKEN, Wako 351-0198, Japan
}

\author{Yong-Hui~Qi}
\email{yonghui.qi@apctp.org}

\affiliation{
Asia Pacific Center for Theoretical Physics, Pohang, Gyeongbuk, 37673, Korea
}

\begin{abstract}
Fractons are particles with restricted mobility.
We give a symmetry-based derivation of effective field theories 
of gapless phases with fractonic topological defects, 
such as solids and supersolids, 
using a coset construction. 
The resulting theory
is identified as the Cosserat elasticity theory, 
which reproduces the conventional symmetric elasticity theory at low energies.
The construction can be viewed as a dynamical realization of the inverse Higgs mechanism. 
We incorporate topological defects such as dislocations and disclinations, 
which are nontrivially related by the Bianchi identities of defect gauge fields. 
The origin of the fractonic nature of defects 
in those systems can be traced back to 
the semidirect product structure of 
translational and rotational groups. 
The construction is immediately extendable 
to higher dimensions and 
systems with broken translational symmetries, 
such as solids, supersolids, and vortex crystals. 
We identify Wess-Zumino terms in supersolids, 
which induce quasiparticle scatterings on 
topological defects. 
\end{abstract}

\maketitle


\section{Introduction}

Fracton phases are a new class 
of quantum phases that host excitations with mobility restrictions~\cite{Nandkishore:2018sel,Pretko:2020cko}. %
Those excitations cannot move at all,  
or 
their motions are restricted in subdimensional spaces. 
Such fracton phases were first discussed in 
exactly solvable lattice models~\cite{PhysRevLett.94.040402,PhysRevA.83.042330,Yoshida:2013sqa,PhysRevB.92.235136,Vijay:2016phm}. 
It has been realized 
that symmetric tensor gauge theories 
\cite{xu2006novel,Xu2006,Pretko:2016kxt, Pretko:2016lgv} 
can encode mobility restrictions 
through the conservation laws of multipole moments~\cite{Pretko:2018qru, Gromov:2018nbv,Bidussi:2021nmp}. 
The elasticity theory of two-dimensional crystals 
was shown to be dual to a symmetric tensor gauge theory, 
and disclinations in solids are fractonic~\cite{Pretko:2018qru}.
Similar dualities can also be formulated for other systems 
such as supersolids~\cite{Pretko:2018tit}
and vortex crystals~\cite{Nguyen:2020yve}.

The relation between the immobility of a particle and 
the conservation of multipole moments
can be seen as follows. 
Suppose that there is a particle number current $j_\mu(x)$, 
which is conserved, $\p^\mu j_\mu =0$. 
If we define the dipole current by 
$(J^a)_\mu (x) = x^a j_\mu$, where $a$ is a spatial index, 
its divergence reads 
\begin{equation}
\p^\mu (J^a)_\mu = j^a. 
\label{eq:dj-j}
\end{equation}
This relation means that the flow of a current 
should be accompanied by the creation/destruction of dipoles. 
Therefore, if the dipole current is conserved, 
$\p^\mu (J^a)_\mu = 0$, 
the particle number current vanishes, $j^a=0$, 
which means that a particle is immobile. 
In the case of a solid in $2+1$ dimensions, 
disclinations and dislocations correspond 
to $j_\mu$ and $(J^a)_\mu$, respectively. 
When the dipole excitations (dislocations) 
are gapped, 
the conservation of dipoles is energetically enforced, 
and disclinations are immobile. 
In this way, the relation~\eqref{eq:dj-j} 
plays a key role in realizing fractons.

In this paper, we discuss 
the construction of the effective field theories of 
gapless phases with fractonic topological defects, 
such as solids and supersolids,
in which the translational symmetry is spontaneously broken. 
We employ a coset construction~\cite{Coleman:1969sm,Callan:1969sn}, 
which is a systematic method of writing down the effective Lagrangian of low-energy theory associated with a spontaneous symmetry breaking (SSB). 
Topological defects appear as a result of an SSB, 
and they can be incorporated in the effective theory. 
In this approach, 
the relations between the defect currents \eqref{eq:dj-j} 
can be traced back to the underlying structure
of the symmetry group and its breaking pattern. 
The geometric origin of the fractonic feature becomes transparent, and 
Eq.~\eqref{eq:dj-j} reflects 
the fact that the rotational symmetry acts nontrivially 
on the translational symmetry. 
Its origin is the same as the Bianchi identities of 
the torsion and curvature of the Riemann-Cartan spacetime. 
The coset construction clarifies this connection.

The rest of the paper is organized as follows. 
In Sec.~\ref{sec:coset},
we give the derivation of the effective theories for solids and supersolids. 
In Sec.~\ref{sec:dual-gauge-theory}, 
we describe the properties of dual gauge theories.
In Sec.~\ref{sec:wz}, we discuss the scattering processes of quasiparticles off topological defects induced by Wess-Zumino terms.  
Section~\ref{sec:summary} is devoted to the summary.

\section{Coset construction and the Cosserat elasticity}\label{sec:coset} 

A coset construction
is a method for constructing the effective theory of 
Nambu-Goldstone (NG) modes associated with an SSB. 
The obtained effective theories are universal, in the sense that their forms are dictated 
by the symmetry-breaking patterns, 
and microscopic details are 
encoded in the values of phenomenological parameters. 
Although it was invented for the breaking of internal symmetries, 
it allows for a number of generalizations, 
such as spacetime symmetries~\cite{Ivanov:1976pg,Low:2001bw,Nicolis:2013lma,Goon:2014ika,Hidaka:2014fra} 
and higher-form symmetries \cite{Hidaka:2020ucc,Landry:2021kko}.

We here perform a coset construction 
for solids and supersolids in $D=d+1$ spacetime dimensions.\footnote{
See Ref.~\cite{Pena-Benitez:2021ipo} for a related construction. 
}
We consider nonrelativistic systems and 
the underlying spacetime symmetry 
is the Galilean group. 
This symmetry is spontaneously broken 
because of the formation of solids or supersolids. 
In the case of  supersolids~\cite{andreev1969quantum}, 
we have an additional $U(1)$ symmetry, 
which is also spontaneously broken. 
In a coset construction, 
we first parametrize the coset space, 
and for each broken generator we have an 
NG field, which is a degree of freedom of the low-energy theory. 
A subtlety for the broken spacetime symmetry is that, 
unlike the case of internal symmetries, 
there is no one-to-one correspondence between 
a broken symmetry generator and a gapless NG mode~\cite{Low:2001bw}. 
When a broken symmetry generator does not commute with 
the translation, 
there can be a nontrivial coupling of NG modes 
through the covariant derivative 
and that results in the reduction of gapless modes 
compared to the number of broken generators. 
For the details of the coset construction 
for broken spacetime symmetries, 
see Ref.~\cite{Nicolis:2013lma}. 

Below, $d$ denotes the spatial dimension, 
and $D = d+1$ is the spacetime dimension. 
We use the mostly-plus convention for 
the Minkowski metric, 
$\eta_{\mu\nu}=(-1,+1,\ldots, +1)$. 
The symmetrization/antisymmetrization 
of indices are denoted by brackets,
$(\cdots)$ and $[\cdots]$, 
respectively. 
For example, 
$A_{(ab)} \equiv (A_{ab} + A_{ba})/2$
and 
$A_{[ab]} \equiv (A_{ab} - A_{ba})/2$. 

\subsection{Symmetry and its breaking pattern}

A Galilean group is the spacetime symmetry group of nonrelativistic systems. 
The generators of the Galilean group in $d$ spatial dimensions, 
$\text{Gal}(d)$, 
satisfy the Galilean algebra $\mathfrak{Gal}(d)$, 
whose nonvanishing commutation relations are 
\begin{equation}
\begin{split}
[J_{ab},J_{cd}]  
&= - 4i\eta_{[a[c} J_{d]b]} ,
\quad
[J_{ab},P_c]  = 2 i \eta_{c [ a} P_{b]}  , \\ 
[J_{ab},B_c] & = 2 i \eta_{c[a} B_{b]} , \quad [B_{a},H]   
 = i P_a, \\
\end{split} ~ \label{Galilei-2}
\end{equation}
where 
$J_{ab}$, $P^a$, and $B^a$
are the generators of rotation, 
spatial translation, 
and Galilean boost, 
respectively, and 
$H$ corresponds to the non-mass energy.  
The Bargmann algebra $\mathfrak{B}(d)$ is a 
central extension of the Galilean algebra $\mathfrak{Gal}(d)$ 
with the following relation, 
\begin{equation}
 [B_{a},P_b] = - i \eta_{ab} Q_0 ,
 \label{eq:bp-q}
\end{equation}
where $Q_0$ represents the total mass, 
which can be written as 
$Q_0 = - mQ$ 
where $m$ is the mass of a particle, and 
$Q$ is the particle number. 
Since $Q$ is a $U(1)$ charge and commutes with every other generator, 
this is a central extension. 
The Bargmann algebra can be obtained 
as the nonrelativistic limit of the Poincar\'{e} algebra \cite{Weinberg:1995mt}. 
For $d \ge 3$, this is the only central extension. 
In the case of two spatial dimensions, 
the Galilean group, $\text{Gal}(2)$, 
admits another central extension. 
In addition to Eq.~\eqref{eq:bp-q}, 
we can have the following nonvanishing commutation relations, 
\begin{equation}
[B_a, B_b] = - i \epsilon_{ab} \kappa \, Q_0 . 
\label{eq:bb-q}
\end{equation}
In addition to $m$, 
there is another parameter $\kappa$, 
which can be interpreted as the spin (per mass) of a particle~\cite{Jackiw:2000tz}.

We consider the following symmetry breaking 
pattern to realize 
solid and supersolid phases, 
\begin{equation}
\text{Gal}(d) 
\times U(1)
\to ({\mathbb Z}^d \rtimes \mathcal G) \times {\mathbb R} , 
\label{eq:ssb-pattern-ss}
\end{equation}
where $\rtimes$ is the semi-direct product 
and $\mathcal G \subset SO(d)$
is a discrete subgroup of $SO(d)$, 
$\mathbb R$ corresponds to the time translation 
and $\mathbb Z^d$ represents discrete spatial translations. 
The low-energy behavior of solids 
can be described by a $d$-dimensional field 
$\phi^a (t, \bm x)$. 
The field $\phi^a (t, \bm x)$ can be regarded as the comoving coordinate of the material. 
The actions of internal rotational 
and translational symmetries are given by 
\begin{equation}
\begin{split}
& \phi^a \mapsto \xi^a_{~b} \phi^b  , \quad \xi^a_{~b} \in SO(d), \\ 
& \phi^a \mapsto \phi^a + c^a, 
\quad c^a \in \mathbb{R}^d . 
\end{split} \label{sym_solid}
\end{equation}
For supersolids, 
to describe the $U(1)$-symmetry breaking, 
we introduce an additional scalar field 
$\phi^0 (t, \bm x)$, 
which is shifted by the $U(1)$ symmetry, 
\begin{equation}
\phi^0 \mapsto \phi^0 + c^0, 
\quad c^0 \in \mathbb R / 2 \pi \mathbb Z \simeq U(1) . 
\end{equation}
The action of the $U(1)$ symmetry 
and the translations can be combined into a four-vector notation, 
\begin{equation}
\phi^A \mapsto \phi^A + c^A , 
\label{sym_supersolid}
\end{equation}
where $c^A$ are constants with index $A=(0,a)$. 
Correspondingly, we introduce a four-vector notation $Q_\mu=(Q_0,Q_a)$. 
When the supersolid (or solid) is in the ground state, 
we can choose the comoving coordinates $\phi^A$ to coincide with the spacetime coordinates $x^A = (x^0, x^a)
$, 
\begin{equation}
\vev{\phi^a} = x^a , 
\quad 
\langle\phi^0 \rangle  = x^0 . 
\end{equation}
We can parametrize the fields 
$\phi^a$ and $\phi^0$ as 
\begin{equation}
\phi^a = x^a + u^a, 
\quad \phi^0 = x^0 + u^0. \label{phi-x-u}
\end{equation}
Here, the field $u^a$ represents the deviation of the material coordinate from its equilibrium position, and is called the displacement field in the elasticity theory. 
The field $u^0$ can be identified as the fluctuating part of the $U(1)$ phase of the condensate (divided by the mass $m$ of a particle).

Let us summarize the unbroken and broken generators for  supersolids: 
\begin{equation}
\begin{split} 
\text{unbroken}: & ~   \bar{P}_\mu = P_\mu + Q_\mu , \quad \text{spacetime \& internal translation} \\ 
& ~  \bar{J}_{ab} = J_{ab} + L_{ab} , ~ \text{spatial \& internal rotation} \\ 
\text{broken}: 
& ~ Q_\mu= (Q_0, Q_a) , \quad \text{internal $U(1) \times \mathbb{R}^{d} $ } \\ 
& ~ L_{ab} , \quad \quad \quad \quad \quad \quad 
\text{internal rotation} \\
& ~  B_a, 
\quad\quad\quad \quad\quad\quad \,
\text{Galilean boost} \\
\end{split} \ii\ii
\end{equation}
Note that the ground state is invariant under the 
combinations of spacetime and internal translations/rotations~\cite{Nicolis:2013lma}. 
The combined generators are written, for translation, as 
$\bar{P}_\mu =  (- H - m Q, P_a + Q_a)$.

\subsection{Covariant derivatives} 

The low-energy degrees of freedom are the coordinate 
of the coset space associated with 
the symmetry breaking~\eqref{eq:ssb-pattern-ss}. 
We call those fields as Nambu-Goldstone (NG) fields. 
We can parametrize the coset space as 
\begin{equation}
\Omega 
 = 
 e^{i x^\mu \bar{P}_\mu  } e^{ i v^a(x) B_a}  e^{ i u^\mu(x) Q_\mu}  e^{ i \theta^{ab}(x) L_{ab}} , 
\label{Omega} 
\end{equation}
where $v^a(x)$, $u^\mu (x)$, and $\theta^{ab}(x)$ are
the NG fields of Galilean boosts, 
$U(1)$ and internal translations, 
and internal rotations, respectively.
The building blocks of the low-energy theory 
can be obtained from the Maurer-Cartan (MC) form, 
\begin{equation}
\omega \equiv -i \Omega^{-1} \rd \Omega . 
\end{equation}
The MC form can be expanded by generators as 
\begin{equation}
\omega = 
\omega_{\bar{P}}^A \bar{P}_A
+\omega_{J}^{ab} \bar{J}_{ab} 
+\omega_Q^\mu Q_\mu 
+\omega_L^{ab} L_{ab} 
+\omega_B^a B_a . 
\end{equation}
The coefficients of the unbroken combined
translation and rotational generator, $\bar P_A$, $\bar{J}_{ab}$ 
give the vielbein and spin connection, 
\begin{equation}
\omega_{\bar{P}}^A = e_\mu^{~A} \rd x^\mu , \quad \omega_{\bar{J}}^{ab} = \omega_\mu^{~ab} \rd x^\mu. 
\end{equation}
The 1-forms proportional to the broken generators 
give the covariant derivatives $D_A$ of NG fields, 
\begin{equation}
\begin{split}
\omega_Q^\nu  
&= e_{\mu}^{~A} D_A u^\nu \rd x^\mu , \\
\omega_L^{ab} 
&= e_{\mu}^{~A} D_A \theta^{ab} \rd x^\mu ,\\
\omega_B^a 
&= e_{\mu}^{~A} D_A v^{a} \rd x^\mu . \\
\end{split}
\end{equation}
By using the commutation relations, we have 
\begin{equation}
\begin{split}
& e^{-i v^c B_c}  P_{a} e^{i v^c B_c} = P_a - v_a Q_0 , \\
& e^{-i v^c B_c} P_0 e^{i v^c B_c} = P_0 -  v^a P_a + \frac{1}{2} v^2 Q_0, \\
& e^{-i\theta^a L_a} \bar{P}_\mu e^{i\theta^a L_a}  = (\bar{P}_0 , P_a +  \xi_a^{~b} Q_b ) ,
\end{split}
\end{equation}
where $\xi_{a}^{~b} \in SO(d)$ 
is the rotational matrix associated with $\theta$, 
$\xi_{a}^{~b} \equiv ( e^{ i \theta \cdot L} )_{a}^{~b}$.
Using those relations, we can identify 
the components of MC 1-forms as 
\begin{equation}
\begin{split}
\omega_{\bar{P}}^0 & = \rd x^0, 
\\
\omega_{\bar{P}}^a &= \rd x^a  - v^a \rd x^0 ,  
\\
\omega_Q^0 & =   \rd u^0 -  v_a \rd x^a +  \frac{1}{2}  v^2 \rd x^0
- \frac{\kappa}{2}v^a \rd v^b\epsilon_{ab}  
, 
\\  
\omega_Q^a & = \rd \phi^b \xi_b^{~a} - \rd x^a + v^a \rd x^0  , \\
\omega_B^a &= \rd v^a . 
\end{split} \label{Mc_1-form_GB}
\end{equation}
The term proportional to $\kappa$ in $\omega_Q^0$ can exist 
only in $(2+1)$-dimensional spacetime. 
The part proportional to the 
rotational generator $L_{ab}$ is written as
\begin{equation}
\omega_L^{ab} L_{ab} = 
- i e^{ - i \theta^{ab} L_{ab}}  \rd e^{ i \theta^{ab} L_{ab}}. 
\end{equation}
From $\omega_{\bar{P}}^A$, 
we can read off the vielbein as
\begin{equation}
e^A = \bigg(\begin{matrix}
\rd x^0  \\
\rd x^a - v^a \rd x^0 \\
\end{matrix} 
\bigg). 
\label{eq:vielbein}
\end{equation}
In terms of components, 
the vielbein and its inverse are given by
\begin{equation}
e_\mu^{~A} 
= \delta_\mu^{~A} - \delta_\mu^{~0} v^A , \quad
e_{~A}^\mu  \equiv (e_\mu^{~A})^{-1} = \delta^\mu_{~A} + \delta^0_{~A} v^\mu , 
\label{e_MA}
\end{equation}
with $v^\mu=(0,v^a)$. 
They satisfy $e^\mu_{~A}e_\mu^{~B} = \delta_A^B$.
We can read off the components of 
$\omega_Q^0, \omega_Q^a,$ and $\omega_B^a$ as 
\color{black} 
\beqa
\begin{split}
(\omega_Q^0)_\mu 
&= 
\partial_\mu  u^0 -  v_a \delta_\mu^{~a} + \frac{1}{2} v^2  \delta_\mu^{~0} 
- \frac{\kappa}{2} \epsilon_{ab} v^a \p_\mu v^b
\color{black}
, \\
(\omega_Q^a)_\mu 
&= 
\partial_\mu \phi^b \xi_b^{~a} - \delta_\mu^{~a} + \delta_\mu^{~0} v^a ,  \\
(\omega_B^a)_\mu 
&= 
\partial_\mu v^a . 
\end{split} \label{deri_1-form}
\eeqa
Accordingly, the covariant derivatives 
of the NG fields are given by 
\beqa
\begin{split}
D_A u^0 
&= 
e_{~A}^\mu 
(\omega_Q^0)_\mu 
=
\partial_A u^0 - \delta_A^{~a} v_a + \delta_A^{~0} \frac{v^2}{2} 
- \frac{\kappa}{2} \epsilon_{ab} v^a \p_A v^b
\color{black}
+ \delta^0_{~A} v^c \left(\partial_c u^0 - v_c - \frac{\kappa}{2} \epsilon_{ab} v^a \partial_c v^b \right) , \\
D_A u^a 
&=
e_{~A}^\mu (\omega_Q^a)_\mu 
= \partial_A \phi^b \xi_b^{~a} - \delta_A^{~a} + \delta_A^{~0} v^a + \delta^0_{~A} v^c (\partial_c \phi^b \xi_b^{~a} - \delta_c^{~a}) , \\
D_A v^a 
&=
e_{~A}^\mu (\omega_B^a)_\mu 
= \partial_A v^a + \delta^0_{~A} v^b \partial_b v^a. \\
\end{split} \ii \label{covdev_ss}
\eeqa
Let us list the temporal and spatial components
of the covariant derivatives separately: 
\begin{equation}
\begin{split}
D_0 u^0 & = \partial_0 u^0 +  v^a \partial_a u^0 - \frac{1}{2}  v^2 
- \frac{\kappa}{2} \epsilon_{ab} v^a \p_0 v^b - \frac{\kappa}{2}\epsilon_{ab}v^a v^c \partial_c v^b
\color{black}
\equiv {\mathcal D}_0 u^0 - \frac{1}{2}  v^2 
- \frac{\kappa}{2} \epsilon_{ab} v^a {\mathcal D}_0 v^b
\color{black}
, \\
D_a u^0 & = \partial_a u^0 - v_a 
- \frac{\kappa}{2} \epsilon_{bc} v^b \p_a v^c
\color{black}
, \\
D_0 u^a 
& = (\partial_0 \phi^b + v^c \partial_c \phi^b)\xi_b^{~a}  \equiv {\mathcal D}_0 \phi^b \xi_b^{~a}, \\
D_b u^a & = \partial_b \phi^c \xi_c^{~a}  - \delta_b^{~a} ,\\
D_0 v^a & =  \partial_0 v^a + v^b \partial_b v^a \equiv {\mathcal D}_0 v^a, \\
D_b v^a  &= \partial_b v^a 
,
\end{split}
\label{eq:Du}
\end{equation}
where we defined 
the convective time derivative, ${\mathcal D}_0$, by\footnote{
Under the action of a Galilean boost, 
\beqa
\begin{split}
 x^a &\mapsto x'^a = x^a + \beta^a t, \\ 
 t   &\mapsto t' = t , \\
 v^a &\mapsto (v^a)' = v^a + \beta^a, \\
\end{split}
\label{galilean_t}
\eeqa
the derivatives are transformed as
$(\partial_0)' = \partial_0 - \beta^a \partial_a$
and $(\partial_a)' = \partial_a$. 
The vielbeins \eqref{eq:vielbein} are transformed covariantly 
under the Galilean transformation, 
$
(\rd x^0)'  = \rd x^0, 
~ (\rd x^a - v^a \rd x^0)' 
 = \rd x^a - v^a \rd x^0 .$
The convective derivative 
${\mathcal D}_0$ is Galilean-invariant, 
$({\mathcal D}_0)' = (\partial_0)' + (v^a)' (\partial_a)' 
 = \partial_0 - \beta^a \partial_a + (v^a + \beta^a) \partial_a = {\mathcal D}_0
$. 
Note that the field $u^0$ is shifted under a Galilean boost 
(in the absence of $\kappa$) as 
$
  (u^0)' 
  = 
  u^0 + \bm \beta \cdot \bm x + \frac{1}2 \beta^2 t . 
$
We can explicitly check that 
the covariant derivative $D_0 u^0$ is indeed invariant under a Galilean boost: 
\begin{equation}
 [ \p_0 u^0 + \frac{1}2 (\p_a u^0)^2 ]'
 = 
 (\p_0 - \bm \beta \cdot \nabla )
 (u^0 + \bm \beta \cdot \bm x + \frac{1}2 \beta^2 t)
 + \frac{1}2 
 (\p_a u^0 + \beta_a)^2
= \p_0 u^0 + \frac{1}{2} (\p_a u^0)^2 . 
\end{equation}
The field $u^a$ is shifted under a Galilean boost as
$(u^a)' = u^a - \beta^a t$ 
and $(\phi^a)'=\phi^a$. 
}
${\mathcal D}_0 \equiv \partial_0 + v^a \partial_a$. 
The covariant derivatives~\eqref{eq:Du} 
are the building blocks of the effective Lagrangian.

\subsection{Effective Lagrangian for supersolids and solids}

For the broken spacetime symmetry, 
not every NG field leads to a physical gapless mode. 
Namely, there can be a redundancy 
in the parametrization of the modes 
and some of them can be expressed in other fields. 
Suppose that the commutator of the unbroken generator $\bar P_\mu$ 
with a broken generator $X$
contains another broken generator $X'$, 
\begin{equation}
    [\bar P_\mu, X] \sim X' + \cdots . 
\end{equation}
If we denote the NG fields for $X$ and $X'$ as 
$\pi$ and $\pi'$, 
we can impose the inverse Higgs constraint (IHC)
of the form $D_\mu \pi' = 0$, 
to express $\pi'$ via the derivative of $\pi$. 
However, whether one {\it should} impose the constraint 
depends on the microscopic details of the system~\cite{Nicolis:2013sga}. 
This choice leads to the differences in the number of gapped physical modes.

Because of the algebra 
$[\bar P_a, B_b] = i \delta_{ab} Q_0$, 
we can express the boost NG field $v^a$ 
in terms of $u^0$. 
By imposing the inverse Higgs constraint 
$D_a u^0 = 0$, 
the velocity field $v^a$ is expressed as (in the absence\footnote{ 
When $\kappa \neq 0$ in $2+1$ dimensions,
the condition $D_a u^0=0$ is written as 
$
    \p_a u^0 - v_a 
    - 
    \frac{\kappa} 2 
    \epsilon_{bc} v^b \p_a v^c  = 0. 
$
If we solve this for $v_a$ perturbatively 
in the number of derivatives, 
\begin{equation}
    v_a = \p_a u^0 
    - 
    \frac{\kappa} 2 
    \epsilon_{bc} \p^b u^0 
    \p_a \p^c u^0 
    + O(\p^5) . 
\end{equation}
Thus, the central extension with $\kappa$ 
gives rise to a correction of the superfluid velocity with higher-order derivatives. 
} of $\kappa$) 
\begin{equation}
v_a = \partial_a u^0 .
\label{eq:ihc-va}
\end{equation}
Namely, we can identify $v_a$ as the superfluid velocity. 
Since $[\bar P_a, L_{bc}] = - 2 i \eta_{a[b} Q_{c]}$, 
it is possible to impose the IHC $D_{[a} u^{b]} = 0$. 
This can be written as $\p_{[b} u^{a]} - \theta^a_{~b} = 0$ to the leading order in fields, 
and by this we can eliminate the antisymmetric part of $\p_{b} u^a$, 
which results in the conventional elasticity theory~\cite{Landau:1986aog,kleinert1989gauge,marsden1994mathematical,chaikin1995principles,zubov1997nonlinear,Beekman:2016szb,Beekman:2017brx} 
written in terms of the symmetric part, $\p_{(b} u^{a)} = (\p_b u^a + \p_a u^b)/2$. 
However, whether one should impose the constraints 
to eliminate certain modes 
depends on the microscopic details of the system~\cite{Nicolis:2013sga}. 
Although the nature of gapless modes does not 
depend on such a choice and hence is universal, 
it can lead to differences in the number of gapped modes.

Here, we {\it choose not to} impose $D_{[a} u^{b]} = 0$, 
and keep the rotational NG field in the Lagrangian. 
We will see that 
this leads to the Cosserat theory of elasticity. 
Substituting Eq.~\eqref{eq:ihc-va} 
to Eq.~\eqref{eq:Du}, the covariant derivatives are 
\beqa
\begin{split}
D_0 u^0 &= 
\partial_0 u^0 + 
\frac{1}{2}   (\partial_a u^0)^2   , \\ 
D_0 u^a  &= ( 
\partial_0 \phi^b  + \partial^c 
u^0 \partial_c \phi^b ) \xi_b^{~a}, \\
D_b u^a & = \partial_b \phi^c \xi_c^{~a}  - \delta_b^{~a} , \\
D \theta^{ab} & = \rd \theta^{ab} + \rd \theta^a_{~c}  \theta^{cb} + O(\theta^3) .
\end{split}
\label{eq:cov-d}
\eeqa
In the absence of rotational NG fields, 
i.e., $\xi_b^{~a}=\delta_b^{~a}$, 
the first two coincide with 
the Galilean invariant 
building blocks of the supersolid effective Lagrangian discussed in Ref.~\cite{Son:2005ak}.

To the leading order in 
the numbers of fields and derivatives, 
the covariant derivatives are written as 
\begin{align}
  D_0 u^0   &\simeq \partial_0 u^0   \\
  D_0 u^a   &\simeq \p_0 u^a   , \\
  D_b u^a   &\simeq \p_b u^a + \theta_b^{~a} ,\\
  D \theta_{ab} &\simeq \rd \theta_{ab} . 
\end{align}
To the quadratic order in the number of fields 
and to the lowest order in derivatives, 
the effective Lagrangian for supersolids 
can be written in the form 
\beqa
\begin{split}
L &= 
\frac{1}2 K (\p_0 u^0)^2 
- \frac{1}2 K_{ij} \p^i u^0 \p^j u^0   
+ 
\frac{1}{2}C_{ab}\partial_0 u^a \partial_0 u^b - \frac{1}{2} C_{iajb} (\partial_i u^a + \theta_i^{~a})(\partial_j u^b + \theta_{j}^{~b}) 
\\
&\quad + \frac{1}{2} C_{iajb}^{(\theta)}\partial_0 \theta_i^{~a}\partial_0 \theta_j^{~b} - \frac{1}{2} C_{iabjcd}^{(\theta)}\partial^i \theta^{ab} \partial^j \theta^{cd}, 
\end{split}
\eeqa
where $C_{iajb}$ are elastic constants. 
There can be an additional term, 
\begin{equation}
{L}^{\text{WZ}} 
= \rho_0 \rd u^0 \wedge \rd u^a \wedge \widetilde{e}_a ,
\label{eq:wz-quadratic}
\end{equation}
where 
$\widetilde{e}_a \equiv \epsilon_{aa_2 \ldots a_{d} }
\rd x^{a_2} \wedge \cdots \wedge 
\rd x^{a_{d}}
$ is a constant $(D-2)$-form.
This term \eqref{eq:wz-quadratic} appears as a Wess-Zumino term, and its derivation is discussed 
in Sec.~\ref{sec:wz}.

The effective Lagrangian for solids can be obtained 
by setting $u^0 = 0$, 
\beqa
\begin{split}
{L} & =  \frac{1}{2}C_{ab}\partial_0 u^a \partial_0 u^b - \frac{1}{2} C_{iajb} (\partial_i u^a + \theta_i^{~a})(\partial_j u^b + \theta_{j}^{~b}) 
+ \frac{1}{2} C_{iajb}^{(\theta)}\partial_0 \theta_i^{~a}\partial_0 \theta_j^{~b} - \frac{1}{2} C_{iabjcd}^{(\theta)}\partial^i \theta^{ab} \partial^j \theta^{cd}.
\end{split}
\eeqa
By choosing the elastic constants as
\begin{equation}
C_{iajb} = C_{ab}\delta_{ij}, 
\quad 
C^{(\theta)}_{iabjcd} = C_{abcd}^{(\theta)}\delta_{ij},    
\end{equation}
the low-energy effective Lagrangian of a solid with translational and rotational NG fields, to the quadratic order, can be written in differential forms as
\begin{equation}
{L} = - C_{ab} \rd_\theta u^a \wedge \star \rd_\theta u^b - C_{abcd}^{(\theta)} \rd \theta^{ab} \wedge \star \rd \theta^{cd} , 
\label{eq:L-cosserat} 
\end{equation}
where 
$\star$ denotes the Hodge dual operation 
and we have defined 
\begin{equation}
\rd_\theta u^a 
\equiv 
\rd u^a - \theta^a_{~b} \rd x^b . 
\label{rd_theta-ua}
\end{equation}
%
%
%
The appearance of the combination \eqref{rd_theta-ua} is anticipated on symmetry grounds. 
The NG fields $u^a$ and $\theta^a_{~b}$ are transformed under infinitesimal translation and rotation as 
\begin{equation}
u^a  \mapsto u^a  - \alpha^a - \beta^a_{~b} x^b , 
\quad \theta^a_{~b}   \mapsto \theta^a_{~b}  - \beta^a_{~b},
\label{u-theta-transfm} 
\end{equation}
where $\alpha^a $ and $\beta^a_{~b}$ are transformation parameters for the translation and rotation, respectively. 
Equation~\eqref{rd_theta-ua} 
is indeed invariant under these transformations. 
Intuitively speaking, 
the reason why the covariant derivative of $u^a$ 
includes $\theta^a{}_b$ is that 
the translational NG field $u^a$ 
is nontrivially transformed under rotations.

The Lagrangian 
\eqref{eq:L-cosserat} 
can be identified with 
that of the Cosserat elasticity theory~\cite{cosserat1909theory,eringen1999theory}.  
The Cosserat theory is a generalization of the conventional elasticity theory, 
and it contains rotational degrees of freedom, 
in addition to the displacement field. 
Such a generalization is necessary 
to explain the properties of 
porous materials~\cite{LAKES198655,YANG198291}
or mechanical metamaterials~\cite{PhysRevLett.120.065501}.
Because of the nontrivial covariant derivative~\eqref{rd_theta-ua}, the rotational NG field $\theta^a{}_b$ has a mass term. 
This introduces characteristic scales 
in the theory, which are absent in the conventional elasticity. 
If we consider the 
long-wavelength limit, 
the mass term is dominant compared 
to the kinetic term 
for $\theta^a_{~b}$, and 
its equation of motion (EOM) is $\theta^a_{~b} = \p_{[b} u^{a]}$.
At sufficiently low energies, 
those gapped modes can be ignored, 
and the remaining part for $\p_b u^a$ is
the symmetric part. 
In this way, the Lagrangian~\eqref{eq:L-cosserat}
reduces to the classical elasticity theory. %
Thus, we have shown that the Cosserat theory arises
naturally from the coset construction, 
and it can be understood 
as a dynamical realization of the inverse Higgs phenomenon~\cite{Ivanov:1975zq,Low:2001bw}.


\subsection{Fractonic topological defects} 

Let us introduce topological defects. 
In a crystal, there are two kinds of defects, 
dislocations and disclinations. 
The former are associated with the translational-symmetry breaking $\mathbb R^d \to \mathbb Z^d$,
and the latter are due to the broken rotational symmetry. 
Dislocations and disclinations can be identified 
with the multivalued~\cite{kleinert2008multivalued}
parts of the fields $u^a$ and $\theta^a_{~b}$, respectively. 
We decompose the covariant derivatives of phonon fields into the continuous part and singular part as 
\begin{equation}
 \rd_{A} u^a  = \rd u^a - \theta^a_{~b} \wedge \rd x^b + A^a  , \quad   
 \rd_{{\mathcal A}} \theta^a_{~b}  = \rd \theta^a_{~b} + {\mathcal A}^a_{~b} , 
 \label{dpi_dlambda}
\end{equation}
where we have introduced 
$A^a \equiv \rd u_{\rm (s)}^a $ and ${\mathcal A}^a_{~b} \equiv \rd \theta_{\rm (s)}^{ab}$
and the subscript ${\rm (s)}$ indicates the singular part.

The way the defect fields, 
$A^a$ and ${\mathcal A}^a_{~b}$, 
enter is the same as that of gauge fields. 
We can promote the translational and rotational 
transformations 
to local ones, 
$\alpha^a \mapsto \alpha^a(x)$ and $\beta^a_{~b} \mapsto \beta^a_{~b}(x)$, 
by simultaneously transforming 
$A^a$ and ${\mathcal A}^a_{~b}$ as 
\beqa
\begin{split}
A^a  \mapsto A^a + \rd\alpha^a + \rd\beta^a_{~b} x^b,  \quad {\mathcal A}^a_{~b}  \mapsto {\mathcal A}^a_{~b} + \rd\beta^a_{~b}. 
\end{split} \label{Aa-Aab}
\eeqa
The gauge-invariant field strengths 
of the defect gauge fields are given by 
\begin{equation}
\star J^a 
\equiv 
\rd A^a + {\mathcal A}_{ab} \wedge \rd x^b  ,\quad 
\star {\mathcal J}^{a}_{~b}
\equiv \rd{\mathcal A}^a_{~b}.  
\label{Ta_Rab}
\end{equation}
We can identify 
$\star J^a$ as the dislocation current, 
and 
$\star {\mathcal J}^{a}_{~b}$ as the disclination current. 
The integration of $\star J^a$ over a surface $S$ 
gives the Burgers vector, $b^a = \int_{S} \star J^a$. 
By taking the exterior derivative of Eq.~\eqref{Ta_Rab}, 
we obtain the following relations, 
\begin{equation}
    \rd \star J^a = \star {\mathcal J}^{ab} \wedge \rd x^b , \quad \rd \star {\mathcal J}^{ab} = 0. \label{dJa-dJab}
\end{equation}
The divergence of the dislocation current is equal to 
the disclination current. 
This relation corresponds to Eq.~\eqref{eq:dj-j} 
and hence represents the fractonic feature of the defects. 
In this derivation, 
the geometric origin~\cite{Gromov:2017vir} 
of the fractonic behavior is manifest: 
the interrelation between the defect currents 
can be traced back to the 
semidirect product structure of 
the translational and rotational groups. 
The origin of those relations 
are the same as the Bianchi identities 
of torsion and curvature in a Riemann-Cartan spacetime~\cite{cartan2001riemannian,Yavari:2012}. 

Let us make several comments. 
We have shown that the inclusion of
rotational NG field, which is a gapped mode, gives rise to the Cosserat elasticity theory, 
and 
the currents of dislocations/disclinations,
which are defects associated with translations/rotations, 
satisfy the Bianchi identities~\eqref{dJa-dJab}. 
This construction depends only on the symmetry-breaking pattern and is generalizable to situations where translations and rotations are spontaneously broken, 
such as supersolids and vortex crystals. 
For each of those systems, 
we can construct a Cosserat-type theory via 
the inclusion of gapped rotational NG fields.
The present derivation is for $D$-dimensional spacetime, 
where $J^a$ and ${\mathcal J}^a_{~b}$ are both
$(D-2)$-form, 
and the dislocations and disclinations are $(D-3)$-dimensional objects. 
They are extended objects in general and 
the motions of disclinations are constrained by 
the relation~\eqref{dJa-dJab}. 
In $3+1$ dimensions, 
those defects are lines~\cite{Pai:2018qnm}.

\section{Dual gauge theories}\label{sec:dual-gauge-theory} 

In this section, 
we study the dual gauge theories 
of the effective theories of 
solids and supersolids 
constructed in the previous section. 
The dual transformations 
of Cosserat elasticity theory 
have been discussed in Refs.~\cite{Gromov:2019waa,Radzihovsky:2019jdo}.

\subsection{Dual gauge theory for solids} 

Let us discuss the properties 
of the dual gauge theory for solids. 
The derivation of the dual theories
is straightforward and 
we give it in Appendix~\ref{app:dual-derivation}. 
An advantage of dualization is that, 
in the dual gauge theory, NG fields couple to the topological defects electrically.\footnote{
Note that we here treat topological defects 
as a background, meaning that they are not dynamical variables in the path integral. 
Incorporating dynamical topological defects 
in three (or higher) spatial dimensions 
in an effective field theory is a nontrivial problem, 
while the particle-vortex duality in 2+1 dimensions is well-established~\cite{Karch:2016sxi, Seiberg:2016gmd}. 
} 
For notational simplicity, 
let us here take the elastic constants to be 
of the form 
$C_{ab} = c \delta_{ab}$, 
$C_{abcd} = c' \delta_{ac} \delta_{bd}$. 
The Lagrangian of the dual gauge theory for solids reads
\begin{equation}
L_{\text{dual}} 
=  - \frac{c}{2}  f^a \wedge \star f_a 
   - \frac{c'}{2}  f^{ab}\wedge \star f_{ab} , 
\qquad  \label{Leff_fa-fab}
\end{equation}
where the gauge-invariant field strengths are defined as 
\begin{equation}
f^a \equiv  \rd a^a   , \quad f^{ab} 
\equiv \rd a^{ab} + \bar c \, \rd x^b \wedge a^a  , 
\label{fa_fab}
\end{equation}
with $\bar c \equiv c / c'$. 
The gauge fields are related to the original fields 
by 
$\rd a^a = \star \rd u^a, \,\,\, \rd a^{ab} = \star \rd \theta^{ab}$. 
In the presence of topological defects, 
we also have source terms, 
$L_{\text{s}} = a^a \wedge \star J_a + a^{ab} \wedge \star {\mathcal J}_{ab}$. 
A dislocation sources translational gauge field $a^a$, 
and a disclination sources rotational gauge field $a^{ab}$. 
The field strengths \eqref{fa_fab} 
are invariant under the following gauge transformations, 
\begin{equation}
a^a  \mapsto  a^a + \rd \lambda^a , 
\quad
a^{ab}  \mapsto  a^{ab} +  \rd \rho^{ab} 
+ \bar c \, \rd x^b \wedge  \lambda^a ,
\end{equation}
where $\lambda^a$ and $\rho^{ab}$ are 
$(D-3)$-form transformation parameters. 
By varying the dual action with respect 
to the gauge fields, 
we obtain a type of Maxwell's equations, 
\begin{equation}
 \rd^\dag f^a  
+ \star^{-1} (\star f^{ab} \wedge \rd x^b) 
= J^a , 
 \quad 
 \rd^\dag  f^{ab} =  {\mathcal J}^{ab}, 
 \label{inhomo}
\end{equation}
where $\rd^\dag$ is the codifferential. 
By applying an exterior derivative on 
$f^a$ and ${f}^{ab}$ in Eq.~(\ref{fa_fab}), we obtain the Bianchi identities, 
\begin{equation}
\rd f^a = 0, \quad \rd {f}^{ab} +\bar c\, \rd x^b \wedge  f^a =0 . \label{homo_Bianchi-f}
\end{equation}
Those equations correspond to the conservation of momentum and angular momentum, respectively.
Equations~(\ref{inhomo}) and (\ref{homo_Bianchi-f}) constitute the set of equations of motion
in the presence of dislocations and disclinations.

To understand its dynamics, 
let us focus on $D=2+1$ and 
rewrite the equations using electric and 
magnetic fields, which are introduced 
for each of $f^a$ and $f^{ab}$ 
by 
\begin{align}
f^a 
&= 
  \epsilon^a{}_b 
  \left( 
  (E^b)_i \rd x^{i} \wedge \rd x^0
  + \frac{1}2 \epsilon_{ij} B^b 
    \rd x^{i}  \wedge \rd x^j 
\right), \\
f^{ab} &= 
    \epsilon^{ab} 
    \left(  
    \mathsf{E}_i  
    \rd x^{i} \wedge \rd x^0 
    + 
    \frac{1}2    \epsilon_{ij} 
    \mathsf{B} \, 
      \rd x^i \wedge \rd x^j 
\right) .
\end{align}
In the absence of topological defects, 
the equations of motion 
are written as follows: 
\begin{itemize}
  \item 
Maxwell's equations for translational phonons: 
\begin{align}
  \p^i (E^a)_i 
  -  \, \mathsf{E}^a 
  &= 0 ,
\\  
  - \tilde \p_i B^a 
+ \p_0 (E^a)_i 
-  \epsilon^{a}{}_i \, \mathsf{B}
&= 0 , 
\end{align}
where $\tilde \p_i \equiv \epsilon_{ij} \p^j$. 
\item 
Bianchi identity for translational phonons: 
\begin{equation}
  -\tilde \p^i (E^a)_i 
  + \p_0 B^a  =0 .
\end{equation}
\item 
Maxwell's equations for rotational phonons: 
\begin{align}
  \p^i \mathsf{E}_i &= 0 , 
   \\
-  \tilde \p_i \mathsf B \,  
+
  \p_0 \mathsf{E}_i 
  &= 0 . 
\end{align}
\item 
Bianchi identity for rotational phonons: 
\begin{align}
  - \tilde \p^i \mathsf{E}_i 
  +  \p_0  \mathsf{B} 
  +  \frac{\bar c} 2  
  \epsilon_c{}^{i}  (E^c)_i 
  &= 0 . 
\end{align}
\end{itemize}
Using the Maxwell's equations and Bianchi identities,\footnote{
See Appendix~\ref{app:dual-solids} 
for more detailed analysis of excitation spectra 
based on the dual gauge theory. 
} 
we can derive closed equations for 
the translational electric fields $(E^a)_i$, 
\begin{equation}
(\p_0)^2 (E^a)_i 
= 
\tilde \p_i  \tilde \p^j (E^a)_j
+ \epsilon^a_{~i} \tilde \p_j \p^k (E^j)_k 
- \frac{\bar c }2 \, \epsilon^a_{~i} 
  \epsilon_c{}^{j}  (E^c)_j ,
\end{equation}
where $\tilde \p_i \equiv \epsilon_{ij} \p^j$ is the derivative in the transverse direction. 
If we see this in the momentum space, 
in the long-wavelength limit, $k \to 0$, 
the antisymmetric part of 
the translational electric fields
$(E^a)_i$ satisfies 
$
\left( 
  \omega^2 - \bar c
\right) \epsilon_a{}^i (E^a)_i 
= 0
$, 
where $\omega$ is the frequency. 
Thus, the antisymmetric part is gapped and its gap 
is given by $\sqrt{\bar c}$. 
At low energies, this part can be dropped, 
and the symmetric tensor gauge theory~\cite{Pretko:2016kxt,Pretko:2016lgv}
is reproduced.

\subsection{Dual gauge theory for supersolids} 

Let us discuss the dual gauge theory 
for supersolids. 
We here start with the following 
quadratic Lagrangian, 
\begin{equation}
L = 
- 
\frac{1}2
C_{ab} 
\rd_\theta u^a 
\wedge \star \rd_\theta u^b
- 
\frac{1}2
C_{abcd} \rd \theta^{ab} \wedge \star \rd \theta^{ab} 
- 
\frac{c_0 }2
\rd u^0 \wedge \star  \rd u^0
- g \, \rd u^0 \wedge \rd u^a \wedge \widetilde{e}_a , 
\label{S-cosserat-0}
\end{equation}
where 
$\widetilde{e}_a \equiv \epsilon_{aa_2 \ldots a_{d} }
\rd x^{a_2} \wedge \cdots \wedge 
\rd x^{a_{d}}
$
is a constant $(D-2)$-form.
The last term arises 
as a Wess-Zumino term, as we discuss later. 
This effective theory contains 
a gapped rotational NG mode $\theta^a_{~b}$, 
and is a Cosserat-type theory for supersolids. 
In the low-energy limit, it reduces 
to the conventional effective theory for supersolids~\cite{Pretko:2018qru}.
The dual gauge theory for supersolids is given by
\begin{equation}
L_{\text{dual}} 
=
- \frac{c_0}{2}  f^0 \wedge \star f^0
- \frac{c}{2} f^a \wedge \star f_a 
- \frac{c'}{2}  f^{ab}\wedge \star f_{ab} 
- g \star f_0 \wedge \star f^a \wedge \widetilde{e}_a . 
\label{eq:ss-dual-l}
\end{equation}
Similarly to the case of solids, 
the field strengths satisfy the following Bianchi identities, 
\begin{equation}
\begin{split}
    \rd f^A &= 0 , \\
    \rd f^{ab} + \bar c \, \rd x^b \wedge f^a &= 0,
\end{split}
\label{eq:dual-bianchi}
\end{equation}
where $A = 0, i$. 
By varying the dual action~\eqref{eq:ss-dual-l} with respect to the gauge fields, 
one obtains the corresponding Maxwell's equations, 
\begin{equation}
\begin{split}
\rd^\dag 
\left( 
f^0 + g_0 \star f^a \wedge \widetilde{e}_a 
\right) 
&= J^0 ,     \\
\rd^\dag
\left( 
f^a -\bar g \star f^0 \wedge \widetilde{e}^{\,a} 
\right) 
+ 
\star^{-1} (\star f^{ab} \wedge \rd x^b) 
&= J^a , 
\\
\rd^\dag  f^{ab} &= {\mathcal J}^{ab}  , 
\end{split}
\label{inhomo-ss}
\end{equation}
where we set 
$g_0 \equiv g / c_0$ and $\bar g \equiv g / c$. 

Similarly to the case of solids, 
let us focus on 2+1 dimensions and 
introduce the electric and magnetic fields for superfluid phonons by 
\begin{equation}
  f^0 
  = 
  e_i \rd x^{i}  \wedge \rd x^0 
  + \frac{1}2 \epsilon_{ij}
  b \, \rd x^{i} \wedge \rd x^j ,
\end{equation}
in addition to the corresponding expressions 
for translational and rotational gauge fields. 
Then, the last term of Eq.~\eqref{eq:ss-dual-l} is 
written as 
\begin{equation}
 -g 
 \left[ 
  b\, (E^a)_{a} 
  + 
  e_a B^a
  \right] 
  \rd x^0 \wedge \rd x^1 \wedge \rd x^2.  
\end{equation}
These are mixed $\bm E \cdot \bm B$-type terms 
and are responsible for a generalized Witten effect~\cite{Pretko:2018tit}. 
Indeed, 
Maxwell's equations for superfluid phonons are written 
in terms of electric and magnetic fields as 
\begin{align}
  \p^i e_i - g_0 \p^i B_i 
  &= -  (J_0)_0 , 
  \label{eq:pe-pb-j0-main}
  \\
  \p_0 e_i - \tilde \p_i b
  + g_0 
  [ \tilde \p_i (E^a)_{a} - \p_0 B_i ]
  &= -  (J_0)_i .
\end{align} 
As pointed out in 
Ref.~\cite{Pretko:2018tit}, 
Eq.~\eqref{eq:pe-pb-j0-main} indicates that 
a vortex acquires a magnetic charge
of the translational gauge field 
because of the topological term, 
which means that a vortex carries 
a crystalline angular momentum. 
For the full set of equations of motion
written by electric and magnetic fields, 
see Appendix~\ref{app:dual-em}.

\section{Wess-Zumino terms for supersolids}\label{sec:wz}

The Lagrangian built from 
the covariant derivatives~\eqref{eq:Du} 
is exactly invariant under the symmetry transformations.
The effective Lagrangian can also 
contain Wess-Zumino terms~\cite{Delacretaz:2014jka,Goon:2012dy}, 
which are invariant only up to total derivatives. 
Such terms arise when the Lie algebra of the symmetry group allows for nontrivial central extensions.
They can be identified 
by finding invariant and closed $(D+1)$-forms
made out of MC forms. 
In this section, we discuss the Wess-Zumino 
terms for supersolids and their phenomenological implications. 

\subsection{Derivation of Wess-Zumino terms} 

In this section, we derive the Wess-Zumino terms 
for supersolids. 
Below, we consider the low-energy limit and do not consider the rotational NG fields $\theta^a_{~b}$. 
We also consider the case where the superfluid density is small,
which is typically the case~\cite{Son:2005ak}. 
We first note that the Maurer-Cartan 
forms satisfy the following equations, 
\beqa
\begin{split}
 \rd \omega_{\bar{P}}^0  &= 0 , 
 \quad 
 \rd \omega_{\bar{P}}^a 
 = \omega_{\bar{P}}^0  \wedge \omega_B^a ,  
 \quad 
 \rd \omega_B^a  = 0 , 
 \\
 \rd \omega_Q^0  &=  
 \omega_{\bar{P}}^a \wedge \omega_B^a - \frac{\kappa}{2} 
 \epsilon_{ab} \omega_B^a \wedge \omega_B^b  , 
 \\ 
 \rd \omega_Q^a  
 &= 
  - \omega_{\bar{P}}^0  \wedge \omega_B^a 
  = - \rd \omega_{\bar{P}}^a, 
\end{split} \label{eq:mc-domega}
\eeqa
where the term proportional to $\kappa$ exists 
only in $2+1$ dimensions.

To look for Wess-Zumino terms, 
we have to find invariant and closed $(D+1)$-forms. 
Let us define the mass current $j_{\rm m}$ 
in $d$-spatial dimensions by 
\begin{equation}
  \star j_{\rm m} 
  \equiv 
  \frac{\rho_0}{d!} 
  \, \epsilon_{a_1 \cdots a_d} 
   \rd \phi^{a_1} \wedge  \cdots \wedge  \rd \phi^{a_d} ,
\end{equation}
where $\rho_0$ is the mass density in the ground state. 
The current is trivially conserved, 
$\rd \star j_{\rm m} = 0$. 
Note that $\rd \phi^a = \omega_{Q}^a + \omega_{\bar P}^a$. 
We have the following closed $(D+1)$-form, 
\begin{equation} 
 \Omega_{D+1} 
= 
 \omega_B^a \wedge (\omega_Q)_a  \wedge 
 \star j_{\rm m}
 = 
- \omega_B^a \wedge  (\omega_{\bar P})_a 
\wedge 
\star j_{\rm m} , 
\end{equation}
where the latter equality follows from 
$\rd \phi^a \wedge \rd \phi^{a_1} \wedge 
\cdots \wedge \rd \phi^{a_d} \epsilon_{a_1 \cdots a_d}
= 0$. 
The closedness of $\Omega_{D+1}$ can be checked 
using Eq.~\eqref{eq:mc-domega}. 
The $(D+1)$-form $\Omega_{D+1}$ can be written as 
\begin{equation}
  \rd v^a  \wedge 
  (\rd x_a - v_a \rd x^0 ) 
  \wedge  \star j_{\rm m}
  = 
  \rd 
  \left[ 
  ( v^a \rd x_a - \frac{v^2}2 \rd x^0  ) 
  \wedge \star j_{\rm m}
  \right] . 
\end{equation}
Therefore, 
we can write down the corresponding Wess-Zumino term 
in $D$ dimensions as 
\begin{equation}
  L^{\rm WZ}_D = 
  ( v^a \rd x_a - \frac{v^2}2 \rd x^0  ) 
  \wedge \star j_{\rm m} . 
\end{equation}
Note that it can be written as 
\begin{equation}
  L^{\rm WZ}_D = 
  (\rd u^0 -\omega_Q^0 )
  \wedge \star j_{\rm m} . 
\label{eq:wz-1}
\end{equation}
The one-form $\omega_Q^0$ is exactly invariant under Galilean boosts. 
Thus, the term proportional to $\omega_Q^0$ 
should be already taken into account by the coset construction. 
Thus, we can adopt the following form instead of Eq.~\eqref{eq:wz-1},\footnote{
The coefficient $\rho_0$ may slightly deviate 
as an effect of a non-zero superfluid fraction. 
}
\begin{equation}
  L^{\rm WZ'}_D = \rd u^0 
  \wedge \star j_{\rm m}. 
\label{eq:wzp}
\end{equation}
This expression is consistent with the interpretation of $\star j_{\rm m}$ as the mass current. 
When the superfluid density is small, 
$\rho_0$ indeed equals the mass density at the equilibrium. 
Hereafter, we consider this situation. 
This term was introduced in Ref.~\cite{Son:2005ak}
for $(3+1)$-dimensional supersolids, 
and it appears as a Wess-Zumino term in the current construction.
It is a total derivative, and 
does not affect the equations of motion. 
Still, this term changes the identification of Noether currents and is needed 
to reproduce the centrally extended algebra, 
$[B_a, P_b] = - i \delta_{ab} Q_0$. 
In the presence of vortices, 
the term induces the interactions of vortices with lattice phonons. 
Under a Galilean boost, $L^{\rm WZ'}_D$ 
is transformed as 
\begin{equation}
  \delta   L^{\rm WZ'}_D = 
  \delta   L^{\rm WZ}_D  = 
  ( \beta^a \rd x_a + \frac{\beta^2}2 \rd x^0 ) 
  \wedge \star j_{\rm m} 
  = 
  \rd 
  \left[
    ( \beta^a  x_a + \frac{\beta^2}2 x^0 ) 
  \wedge \star j_{\rm m} 
  \right], 
\label{eq:wz-tr}
\end{equation}
which is a total derivative.

In 2+1 dimensions, 
the Galilean algebra allows for another central extension parametrized by $\kappa$. 
Correspondingly, we have another invariant 4-form, 
\begin{equation}
\Omega'_{4}
\equiv   
\frac{\kappa} 2
\epsilon_{ab} \, \omega_B^a \wedge \omega_B^b 
\wedge \star j_{\rm m}. 
\end{equation}
It produces the following Wess-Zumino term, 
\begin{equation}
{L}^{\rm WZ2}_{D=3}
= 
\frac{\kappa} 2
\epsilon_{ab} v^a \rd v^b 
\wedge \star j_{\rm m} . 
\label{eq:wz-2}
\end{equation}
Under a Galilean boost, this term is shifted by a total derivative, 
\begin{equation}
 \delta 
{L}^{\rm WZ2}_{D=3}
= 
  \frac{\kappa} 2
  \epsilon_{ab} \beta^a \rd v^b 
  \wedge \star j_{\rm m}
=   
\rd
\left[ 
  \frac{\kappa} 2
  \epsilon_{ab} \beta^a  v^b 
  \wedge \star j_{\rm m}
\right] 
\end{equation}
If we use the IHC $v_a = \p_a u^0$, the term \eqref{eq:wz-2} also is a total derivative, 
and the EOMs are not affected in the absence of topological defects. 

Wess-Zumino terms change 
the identification of the currents. 
The term \eqref{eq:wzp} leads to the following additional contribution to the boost current,  
\begin{equation}
    (\star j_{B}^a)_{\rm WZ} =
    \frac{\p L^{\rm WZ'}_D}{\p \rd u^0} \delta^a u^0 
    = 
    x^a (\star j_{\rm m}) ,
\end{equation}
where $\delta^a u^0$ indicates the variation of $u^0$ 
under a Galilean boost in the $a$-th direction. 
The boost current is written as 
\begin{equation}
    \star j_{B}^a 
    = - t (\star p^a) + x^a (\star j_{\rm m}) ,
\end{equation}
where $p^a$ denotes the translational current. 
With this contribution, 
we can reproduce the centrally extended algebra, 
\begin{equation}
    \langle 
    [P_a, B^b ]
    \rangle 
=
    \langle 
    [P_a, \int_V \star j_B^b]
    \rangle 
    = 
    \langle 
     \int_V (-i \p_a  (x^b \star j_{\rm m})) 
    \rangle 
    = 
    - i \delta_{a}^{~b}
    \langle 
     \int_V  \star j_{\rm m}
     \rangle 
    =
    - i \delta_{a}^{~b} \langle Q_0 \rangle . 
\end{equation}

\subsection{Scattering of quasiparticles off topological defects}

The Wess-Zumino terms 
\eqref{eq:wzp} and \eqref{eq:wz-2}
are total derivatives and 
do not affect the EOM in the absence of topological defects. 
However, they change the identification of Noether currents, 
and when topological defects are present, 
these terms induce nontrivial scattering effects.\footnote{
The elastic scattering of lattice phonons 
off superfluid vortices through Eq.~\eqref{eq:wzp} 
is studied in Ref.~\cite{Son:2005ak}. } 
Here, we discuss such processes induced by 
Eqs.~\eqref{eq:wzp} and \eqref{eq:wz-2}. 
Since their coefficients are determined by the symmetry algebra, 
the coupling constants of those processes 
are model-independent.

Let us first discuss the consequence of Eq.~\eqref{eq:wzp}. 
We here consider a vortex in a 
$3+1$-dimensional supersolid 
located along the $z$ direction at $x=y=0$. 
Such a vortex configuration can be expressed by 
$\epsilon_{abc }\p^a \p^b u^0 = 
(2\pi/m) \delta_c (\bm x_{\rm T})$, 
where 
$\delta_c (\bm x_{\rm T}) \equiv n_c \delta (x) \delta (y)$ 
is the transverse delta function 
and 
$\bm n = \hat{\bm z}$ is the unit tangent vector along the vortex. 
In the presence of a superfluid vortex, 
the Lagrangian density can be written as 
\beqa
\begin{split}
L^{\rm WZ'}_{D=4} 
& = - \frac{\rho_0}{ 3! m} \epsilon^{\mu\nu\rho\sigma} \epsilon_{abc} \partial_\mu \partial_\nu \varphi \phi^a \partial_\rho \phi^b \partial_\sigma \phi^c \\
&=  - \pi \frac{\rho_0}{m} 
\epsilon_{abc} u^a \dot{u}^b \delta^c(\bm x_{\rm T}) 
+ O(u^3). 
\end{split}
\eeqa
The leading-order term 
describes elastic scatterings of lattice phonons 
off superfluid vortices in a supersolid.
We denote the energy/momentum 
and polarization of the incoming/scattered 
particles as 
$\{ (\omega, \bm k), \bm \epsilon \}$
and 
$\{ (\omega', \bm k'), \bm \epsilon' \}$, 
respectively. 
For definiteness, 
we consider the incoming lattice phonon
that is incident perpendicularly to the vortex, 
$\bm k \cdot \bm n =0$, 
and assume that it is transversely polarized, 
$\bm k \cdot \bm \epsilon = 0$. 
We also assume that the 
scattered phonon is also transversely polarized, 
and the polarization vectors of 
the initial and scattered phonons 
are in the plane perpendicular to the vortex, 
$\bm \epsilon \cdot \bm n = \bm \epsilon' \cdot \bm n = 0$. %
The current collision geometry reduces 
the problem effectively to two-spatial dimensions. 
The matrix element of this 
elastic scattering process is\footnote{
The scattering processes involving vortices 
have been studied in the EFT approach, 
for example, in Refs.~\cite{Endlich:2010hf,Nicolis:2011cs,Horn:2015zna}. 
In the computation of scattering amplitudes, 
we need to take into account the noncanonical normalization of the kinetic term~\cite{Endlich:2010hf}, 
\begin{equation}
{\mathcal L} = \frac{\rho_0}{2}[(\dot{u}^a)^2 - v^2 (\partial_i u^a)^2] 
+ \frac{\rho_s}{2} (\dot{u}^0)^2 + \ldots . 
\end{equation}
}
\begin{equation}
i {\mathcal M} 
= 
2 \cdot 
\frac{1}{\rho_0}\frac{\pi \rho_0}{m} i \omega \epsilon_{ab c} 
\epsilon^a \epsilon^b n^c
= 
\frac{2 \pi }{m} 
\omega 
(\bm \epsilon \times \bm \epsilon')
\cdot \bm n
= 
\frac{2 \pi }{m} 
\omega \sin \theta ,
\end{equation}
where $\theta$ is the angle between $\bm k$ and $\bm k'$. 
We here assume that the supersolid is isotropic. 
The $1$-body final-state phase space is given by 
\begin{equation}
\rd \Pi_b 
\equiv 2\pi \delta(\omega - \omega') \frac{1}{2\omega_b}  \frac{\rd^2k_b}{(2\pi)^2} .
\end{equation}
Since the superfluid vortex does not break time translations, 
the scattering conserves the energy, $\omega=\omega'$, 
which is reflected in 
the phase-space delta function, $2\pi \delta(\omega-\omega')$. 
The infinitesimal cross section 
of a vortex line element $\rd \ell$ is given by 
\beqa
\begin{split}
\rd \sigma_{a\to b} & = \frac{1}{2 \omega } 
\frac{1}{v_a } |{\mathcal M}_{a\to b}|^2 \rd \Pi_{b}  \rd \ell ,  
\\
& = 
\frac{1}{2 \omega }  
\frac{1}{v_{\rm T}}  
\left( \frac{4 \pi^2 }{m^2} \omega^2  \sin^2\theta  \right) 
\left( 
\frac{\rd \theta  \rd \ell }{4\pi} \frac{1}{v_{\rm T}^2}  \right) 
\\
& = 
\frac{\pi}{2 }  \frac{\omega }{m^2 v_{\rm T}^3} 
\sin^2\theta  \rd \theta \rd\ell   , 
\end{split}
\eeqa
where $v_{\rm T}$ is the velocity of transverse phonons. 
The differential cross section per unit vortex length is written as 
\begin{equation}
\frac{\rd^2 \sigma }{\rd \theta \rd \ell } 
= 
\frac{\pi}{2} 
\frac{k}{m^2 v_{\rm T}^2}  \sin^2\theta , 
\end{equation}
which is obtained in Ref.~\cite{Son:2005ak}. 
The cross section is linearly proportional 
to the momentum $k$ and
it is largest when the scattering angle is $\pi/2$.
This is the dominant elastic scattering process 
of lattice phonons off vortices 
when the superfluid density is small.

Let us now consider 
a dislocation in a $(3+1)$-dimensional supersolid. 
The term $\rd u^0 \wedge \star j_{\rm m}$ 
leads to the conversion of lattice phonons and superfluid phonons
on a dislocation. 
The existence of a dislocation leads to the following 
multivalued part of $u^a$, 
\begin{equation}
\epsilon_{bcd}\partial^b \partial^c u^a 
= b^a \delta_d (\bm x_{\rm T}), 
\end{equation}
where $b^a$ is the Burgers vector. 
The interaction Lagrangian is written as 
\beqa
\begin{split}
L^{\rm WZ'}_{D=4} 
& = \frac{\rho_0}{3!} \rd u^0\wedge  \rd \phi^a \wedge \rd \phi^b \wedge \rd \phi^c \epsilon_{abc} \\
& = - \frac{\rho_0}{2} u^0\wedge  \rd^2 \phi^a \wedge \rd \phi^b \wedge \rd \phi^c \epsilon_{abc}\\
&= 
- \rho_0 u^0  
\epsilon_{abc}
b^a \dot{u}^b \delta^c(\bm x_{\rm T})
\rd x^0 \wedge \rd x^1 \wedge \rd x^2 \wedge \rd x^3 
+ O(u^3) . 
\end{split}
\eeqa
This results in the conservation of a 
superfluid phonon 
with energy/momentum $(\omega, \bm k)$ 
to a lattice phonon $(\omega', \bm k')$
with polarization $\bm \epsilon'$ 
via the scattering off a dislocation. 
The matrix element of this process is
\begin{equation}
i {\mathcal M} 
= 
\frac{1}{\sqrt{\rho_s} \sqrt{\rho_0 } } 
\rho_0 
i \omega' b^a \epsilon'^b \epsilon_{abc}  n^c 
= 
i 
\sqrt\frac{\rho_0}{ \rho_s  }
\omega' 
(\bm b \times \bm \epsilon' ) \cdot \bm n ,
\end{equation}
where $\bm n$ is the tangent vector to the dislocation. 
The infinitesimal cross section is 
\beqa
\begin{split}
\rd \sigma & = \frac{1}{2 \omega }  \frac{1}{v_s} 
|{\mathcal M}|^2
\left( \frac{ \rd \theta \rd \ell}{4\pi}  \frac{1}{v^2} \right) 
\\
& =  
\frac{\rho_0}{\rho_s} 
\frac{ k }{8 \pi v^2} 
|(\bm b \times \bm \epsilon' ) \cdot \bm n |^2 
\rd\theta \rd \ell ,
\end{split}
\eeqa
where $v_s$ and $v$ are the velocities of 
superfluid phonons and lattice phonons, 
respectively, 
and $\theta$ is the angle between $\bm k$ and $\bm k'$. 
The cross section is determined by the relative 
orientations of the Burgers vector $\bm b$, 
the polarization $\bm \epsilon'$ of the final-state lattice phonon, 
and the direction $\bm n$ of the dislocation. 
For example, the cross section vanishes 
for a screw dislocation, for which 
$\bm b$ is parallel to $\bm n$. 
We also note that the cross section is enhanced 
at small superfluid densities.

Next, let us consider the term~\eqref{eq:wz-2},
which can exist in 2+1 dimensions. 
In the presence of vortices, this term induces 
nontrivial interactions among NG fields. 
In 2+1 dimensions, a vortex is a point-like object. 
If we consider a vortex placed at the origin, 
we have $\epsilon_{ab} \p^a \p^b u^0 
=\frac{ 2 \pi}{m} \delta (\bm x)$, 
where 
$\delta (\bm x) \equiv \delta (x) \delta (y)$, 
and we have the following terms, 
\begin{equation}
  \frac{2\pi \kappa \rho_0} m \delta (\bm x) 
  \left( 
    \dot u^0 \, \nabla \cdot \bm u
    - \dot u^a \p_a u^0 
    + O(u^3)
    \right) . 
\end{equation}
This indicates that superfluid phonons 
and lattice phonons can be converted 
via the scatterings off a superfluid vortex. 
Let us consider the process 
of a transverse lattice phonon with $(\omega, \bm k)$ 
and polarization $\bm \epsilon$ 
converted on a vortex into 
a superfluid phonon with $(\omega', \bm k')$. 
The matrix element is 
\begin{equation}
i {\mathcal M} 
=
\frac{1}{\sqrt{\rho_0 \rho_s}}
\frac{2\pi \kappa \rho_0}{m}
\left[ 
- (- i \omega) (- i \bm k') \cdot \bm \epsilon 
\right]  
=
\frac{2 \pi \kappa}{m} \sqrt{\frac{\rho_0}{\rho_s}}
\omega
\bm k' \cdot \bm \epsilon . 
\end{equation}
The infinitesimal cross section is computed as 
\begin{equation}
\begin{split}
\rd \sigma & = 
\frac{1}{2 \omega }  
\frac{1}{ v } 
|{\mathcal M}|^2 \rd\Pi_{b} , 
\\
& = \frac{1}{2 \omega } 
\frac{1}{v} 
\left(  \frac{4 \pi^2 \kappa^2}{m^2} \frac{\rho_0}{\rho_s} \omega^2 
(\bm k' \cdot \bm \epsilon)^2 
\right) 
\left( \frac{ \rd\theta}{4\pi}  \frac{1}{v_s^2} \right)  
\\
&= 
\frac{\pi}{2} \frac{\rho_0}{\rho_s}
\frac{\kappa^2 }{m^2  v_s^2}k  k'^2 \sin^2\theta  
\rd \theta , 
\end{split}
\end{equation}
where $\theta$ is the angle between $\bm k$ and $\bm k'$. 

Finally, let us emphasize that the coupling constants associated with the Wess-Zumino terms are fixed by the symmetry algebra, 
and the cross sections associated with those terms 
are model independent predictions of the effective theory.

\section{Summary}\label{sec:summary}

We derived effective field theories 
of gapless phases with fractons,
such as solids and supersolids, using a coset construction. 
We found that a dynamical realization of the inverse Higgs phenomenon 
naturally leads to the Cosserat theory of elasticity. 
The topological defects appear 
as the singular parts of the NG fields, 
and the corresponding currents obey 
the relation~\eqref{dJa-dJab}, 
which plays a key role so that the disclinations behave as fractons. 
The derivation clarifies the geometric origin of 
the fractonic nature: 
it comes from the semidirect product structure 
of the translational and rotational groups. 
The current construction can be applied 
to systems where the translational symmetry is broken, 
and we can understand why fractons appear in 
solids, supersolids, vortex crystals, and so on. 
We identified Wess-Zumino terms in supersolids, 
which differ by total derivatives 
under symmetry transformation. 
Those terms 
induce nontrivial scattering processes 
involving topological defects 
in a supersolid phase. 
We gave examples of the computations of 
scattering cross sections of such processes. 
When the superfluid density is small, 
the coupling constants of these processes 
are fixed by the algebra, and hence are model independent.

\begin{acknowledgments}
We thank Gil-Young Cho and Ki-Seok Kim for valuable discussions. 
Y.~H. and Y.~H.~Q. are supported by the Korean Ministry of Education, Science, and Technology, Gyeongsangbuk-do Provincial Government, and Pohang City Government for Independent Junior Research Groups at the Asia Pacific Center for Theoretical Physics (APCTP). 
Y.~H. is supported by the National Research Foundation (NRF) funded by the Ministry of Science of Korea (Grant No. 2020R1F1A1076267).
\end{acknowledgments}

\appendix

\section{Details on dual gauge theories}\label{app:dual} 

In this appendix, we provide the derivation of dual gauge theories 
of the effective Lagrangians of solids and supersolids. 
We also derive the corresponding Maxwell's equations 
written in terms of electric and magnetic fields, 
and discuss their dynamical properties.

\subsection{Derivation of dual gauge theories}\label{app:dual-derivation}

In this section, we give the derivation of the dual gauge theories for solids and supersolids. 
Since the gauge theory of solids can be obtained 
from that of supersolids, we here discuss supersolids. 
We start with the partition function of supersolids 
$Z= \int [{\mathcal D}u_a] [{\mathcal D} \theta_{ab}]  e^{i S[u^a, \theta^{ab}, u^0] }$, 
where the action is given by 
\begin{equation}
S[u^0, u^a, \theta^{ab}] 
= - \int_{{\mathcal M}_D} 
\left[
\frac{1}2
C_{ab} 
\rd_\theta u^a 
\wedge \star \rd_\theta u^b
+
\frac{1}2
C_{abcd} \rd \theta^{ab} \wedge \star \rd \theta^{ab} 
+ 
\frac{1}2
C_{00} \rd u^0 \wedge \star  \rd u^0
+ g \, \rd u^0 \wedge \rd u^a \wedge \widetilde{e}_a 
\right]
, 
\label{S-cosserat-0} 
\end{equation}
where ${\mathcal M}_D$ is 
a $D$-dimensional spacetime manifold.
For notational simplicity, 
we organize $u^0$ and $u^a$ as a four-vector 
as $u^A =(u^0, u^a)$, 
and the constants are also organized accordingly. 
We write the action as 
\begin{equation}
S[u^A, \theta^{ab}] 
= - \int_{{\mathcal M}_D} 
\left[ 
\frac{1}2
C_{AB} 
\rd_\theta u^A
\wedge \star \rd_\theta u^B
+ 
\frac{1}2
C_{abcd} \rd \theta^{ab} \wedge \star \rd \theta^{ab} 
+ g \, \rd u^0 \wedge \rd u^a \wedge \widetilde{e}_a 
\right] 
. 
\label{S-cosserat-0} 
\end{equation}
By introducing auxiliary fields, 
$(\tau^A, \sigma^{ab})$, 
the partition function can be written as 
$Z= \int [{\mathcal D}u^A]
[{\mathcal D} \theta^{ab}]  [{\mathcal D}\tau^A] [{\mathcal D}\sigma^{ab}] e^{i S[u^A, \theta^{ab}, \tau^A, \sigma^{ab}] }$, 
where the action is given by  
\begin{equation}
 S 
 =
 \int_{ {\mathcal M}_D}  
 \left[ 
 \frac{1}2 
   C^{-1}_{AB}
 \tau^A \wedge \star \tau^B 
 + \tau^A \wedge \star \rd_{\theta} u_A 
+
\frac{1}2 
C^{-1}_{abcd} \sigma^{ab} \wedge \star \sigma^{cd} + \sigma^{ab} \wedge \star \rd \theta_{ab} 
- g \, \rd u^0 \wedge \rd u^a \wedge \widetilde{e}_a 
\right] 
.
\label{Z-S_u-A_theta-A_tau-sigma}
\end{equation}
By doing a variation with respect to $\tau^a$ and $\sigma^{ab}$, we obtain 
\begin{equation}
- C^{-1}_{AB} \tau_B 
 = \rd_{\theta} u^A, 
\quad 
- C^{-1}_{abcd} \sigma_{cd}  
= \rd \theta^{ab} . 
\label{eom-tau-u_sigma-omega}    
\end{equation}
The path-integration of (the smooth part of) 
$u^A$ and $\theta^{ab}$ leads to the following 
equations of motion, 
\begin{equation}
\rd \star  \tau^A =0 , 
 \quad  \rd \star  \sigma^{ab}  +   \rd x^b \wedge \star \tau^a   = 0, 
 \label{continu-eom}    
\end{equation}
which are conservation laws of 
$U(1)$ charge $(\tau^0)_0$, 
momentum density 
$(\tau^a)_0$, 
and angular momentum density$(\sigma^{ab})_0$, respectively. 
The equations above can be solved explicitly 
by introducing $(D-2)$-form gauge fields 
$a^A, a^{ab}$ as 
\begin{equation}
\star \tau_A  =  -  C_{AB} \rd a^B, 
\quad 
\star  \sigma_{ab}  
=  - C_{abcd} \rd a^{cd} -  C_{ac}  \rd x_b \wedge a^c. 
\label{eq:tau-sigma-a}
\end{equation}
By comparing with Eq.~(\ref{eom-tau-u_sigma-omega}), 
we obtain the relation between the dual gauge fields 
and the original fields as 
\begin{equation}
\rd a^A = \star \rd u^A, \quad \rd a^{ab} = \star \rd \theta^{ab}.
\end{equation}
Substituting Eq.~\eqref{eq:tau-sigma-a} to 
Eq.~\eqref{Z-S_u-A_theta-A_tau-sigma}
gives the following dual effective Lagrangian, 
\begin{equation}
L_{\text{dual}} 
= 
-\frac{1}{2} C_{AB} f^A \wedge \star f^B 
-\frac{1}{2} C_{abcd} f^{ab}\wedge \star f^{cd} 
- g \star f_0 \wedge \star f^a \wedge \widetilde{e}_a . 
\label{Leff_fa-fab}
\end{equation}
To simplify the expression, 
let us choose isotropic elastic constants as 
$C_{ab} = c \delta_{ab}$ and 
$C_{abcd} =c' \delta_{ac}\delta_{bd}$, 
and $C_{00} = c_0$. 
Then, we have 
$\star \tau_a = -  \rd a_a$, 
and 
$\star  \sigma_{ab}  = - \rd a_{ab} - \bar c \, \rd x_b \wedge  a_a $
with $\bar c \equiv c/c'$. 
In this case, the dual Lagrangian reads
\begin{equation}
\begin{split}
L_{\text{dual}} 
& =
- \frac{c_0}{2} 
f^0 \wedge \star f^0 
- \frac{c}{2} 
f^a \wedge \star f_a 
- \frac{c'}{2} f^{ab}\wedge \star f_{ab} 
- g \star f_0 \wedge \star f^a \wedge \widetilde{e}_a 
, \\
\end{split} \qquad  \label{Leff_fa-fab}
\end{equation}
where the field strengths are defined as\footnote{
In the general case, $f^a\equiv \rd a^a$ and $f^{ab} \equiv \rd a^{ab} + (C^{-1})^{cdab} C_{ce} \, \rd x_d \wedge a^e$. 
}
\begin{equation}
f^A \equiv  \rd a^A  , \quad f^{ab} 
\equiv \rd a^{ab} + \bar c \, \rd x^b \wedge a^a  . 
 \label{app:fa_fab}
\end{equation}
The field strengths \eqref{app:fa_fab} 
are invariant under the following gauge transformations, 
\begin{equation}
a^A \mapsto  a^A + \rd \lambda^A , 
\quad
a^{ab}  \mapsto  a^{ab} + \rd \rho^{ab} 
+ \bar c \, \rd x^b \wedge  \lambda^a ,
\end{equation}
where $\lambda^A$ and $\rho^{ab}$ are 
$(D-3)$-form gauge parameters.
Note that the rotational gauge field is shifted 
by the transformation parameter of the translational gauge field. 
Accordingly, the field strengths satisfy the following 
Bianchi identities, 
\begin{equation}
\begin{split}
    \rd f^A &= 0 , \\
    \rd f^{ab} + \bar c \, \rd x^b \wedge f^a &= 0. 
\end{split}
\label{eq:dual-bianchi}
\end{equation}
Topological defects enter as sources to the dual gauge fields, $L_{\text{s}} = C_{AB} a^A \wedge \star J^B + 
C_{abcd} a^{ab} \wedge \star {\mathcal J}^{cd}$. 
By varying the dual action with respect to the gauge field $a^A$ and $a^{ab}$, we obtain the equations of motion as 
\begin{equation}
\begin{split}
\rd^\dag 
\left( 
f^0 + g_0 \star f^a \wedge \widetilde{e}_a 
\right) 
&= J^0 ,     \\
\rd^\dag
\left( 
f^a -\bar g \star f^0 \wedge \widetilde{e}_a 
\right) 
+ 
\star^{-1} (\star f^{ab} \wedge \rd x^b) 
&= J^a , 
\\
\rd^\dag  f^{ab} &= {\mathcal J}^{ab}  , 
\end{split}
\label{app:inhomo}
\end{equation}
where we set 
$g_0 \equiv g / c_0$ and $\bar g \equiv g / c$. 
The equations of motion~\eqref{app:inhomo} 
together with 
Bianchi identities \eqref{eq:dual-bianchi} 
describe the dynamics of 
superfluid phonons, 
translational phonons, 
and rotational phonons 
in $D$ spacetime dimensions 
in the presence of vortices, dislocations, 
and disclinations. 

\subsection{Equations of motion in terms of electric and magnetic fields}\label{app:dual-em}

Let us rewrite the EOMs using electric and magnetic fields. 
We here consider 2+1 dimensions. 
Electric and magnetic fields 
for superfluid phonons, lattice phonons, and rotational 
phonons are introduced by 
\begin{align}
  f^0 
  &= 
  e_i \rd x^{i} \wedge \rd x^{0}
  + \frac{1}2 \epsilon_{ij}
  b \, \rd x^{i} \wedge \rd x^{j}, 
  \\
  f^a 
  &= 
  \epsilon^a{}_b 
  \left(
  (E^b)_i \rd x^{i} \wedge \rd x^{0}  
  + \frac{1}2 \epsilon_{ij} B^b 
     \rd x^{i} \wedge \rd x^{j} 
\right) ,  
  \\
   f^{ab} 
    &= 
    \epsilon^{ab} 
    \left(
    \mathsf{E}_i
    \rd x^{i} \wedge \rd x^{0}  
    + 
    \frac{1}2    \epsilon_{ij} 
    \mathsf B
    \, 
   \rd x^{i} \wedge \rd x^{j} 
   \right) . 
\end{align}
The topological term can be written as 
\begin{equation}
-g  \star f_0 \wedge \star f^a \wedge \widetilde e_a 
 = 
 -g 
 \left[ 
  b\, (E^a)_{a} 
  + 
  e_a B^a
  \right]
  \rd x^0 \wedge \rd x^1 \wedge \rd x^2 . 
\end{equation}
Those are crossed $\bm E \cdot \bm B$-type 
terms and are responsible for the generalized Witten effect. 

Maxwell's equations for superfluid phonons 
are now written as 
\begin{align}
  \p^i e_i - g_0 \p^i B_i 
  &= -  (J_0)_0 , 
  \label{eq:pe-pb-j0}
  \\
  \p_0 e_i - \tilde \p_i b
  + g_0 
  [ \tilde \p_i (E^a)_{a} - \p_0 B_i ]
  &= -  (J_0)_i ,
\end{align} 
where $\tilde \p_i \equiv \epsilon_{ij} \p^j$. 

Maxwell's equations for translational phonons are 
\begin{align}
  \p^i (E^a)_i 
  +\bar g \p^a b 
  -  \, 
  \mathsf{E}^a
  &=  \epsilon^{a}_{~b} (J^b)_0  , 
\\  
  - \tilde \p_i B^a 
+ \p_0 (E^a)_i 
+ 
\bar g 
\left( 
  \delta^{a}_{~i} \, \p_0 b 
-  
 \tilde \p_i e^a 
\right) 
-  \,  \epsilon_{a i} 
\mathsf{B} 
&= \epsilon^{a}_{~b} (J^b)_i . 
\end{align}

Maxwell's equations for rotational phonons are 
\begin{align}
  \p^i \mathsf{E}_i &= 
  -  \frac{1}2  \epsilon_{ab} (\mathcal J^{ab})_0  , 
  \\
  - \tilde \p_i \mathsf{B} 
  \,  
   +
  \p_0 \mathsf{E}_i
  &= 
  -  \frac{1}2  \epsilon_{ab} (\mathcal J^{ab} )_i . 
\end{align}

Bianchi identities are written as 
\begin{align}
  - \tilde \p^i  e_i 
  + \p_0 b &= 0 ,\\
  - \tilde \p^i (E^a)_i + \p_0 B^a  &=0 , 
  \\
  - \tilde \p^i \mathsf{E}_i 
  +  \p_0  \mathsf{B} 
  +  \frac{\bar c } 2  \epsilon_c{}^{i}  (E^c)_i 
  &= 0 . 
\end{align}

\subsection{Excitations in solids}\label{app:dual-solids}

Let us look at the properties of excitations in solids
in terms of the dual gauge theory. 
The EOMs for solids 
can be obtained by setting $f^0 = 0$. 
In the absence of topological defects, 
The EOMs are written as follows: 
\begin{itemize}
  \item 
Maxwell's equations for translational phonons: 
\begin{align}
  \p^i (E^a)_i 
  -   \mathsf{E}^a 
  &= 0 ,
\\  
  - \tilde \p_i B^a 
+ \p_0 (E^a)_i 
-  \epsilon^{a}{}_i \, \mathsf{B}
&= 0 , 
\end{align}
\item 
Bianchi identity for translational phonons: 
\begin{equation}
  -\tilde \p^i (E^a)_i 
  + \p_0 B^a  =0 , 
\end{equation}
\item 
Maxwell's equations for rotational phonons: 
\begin{align}
  \p^i \mathsf{E}_i &= 0 , 
   \\
- \tilde \p_i \mathsf{B} \,  
+ \p_0 \mathsf{E}_i 
  &= 0 . 
\end{align}
\item 
Bianchi identity for rotational phonons: 
\begin{align}
  - \tilde \p^i \mathsf{E}_i 
  +  \p_0  \mathsf{B} 
  +  \frac{\bar c} 2  
  \epsilon_c{}^{i}  (E^c)_i 
  &= 0 . 
\end{align}
\end{itemize}

Let us write the EOM in term of electric fields. 
The translational electric fields 
$(E^a)_i$ have four components. 
Because of the Gauss law, 
$\p_a \p^i (E^a)_i =0$, 
there are three physical degrees of freedom. 
The EOM for $(E^a)_i$ can be derived as 
\begin{equation}
\begin{split}
(\p_0)^2 (E^a)_i 
&= 
 \tilde \p_i \p_0 B^a 
+  \epsilon^{a}_{~i} \p_0 \mathsf{B}
\\
&=     
 \tilde \p_i  \tilde \p^j (E^a)_j
+  \epsilon^{a}_{~i}
[
   \tilde \p^j \mathsf{E}_j 
  -  \frac{\bar c} 2  \epsilon_c{}^{j}  (E^c)_j 
]\\
&=     
 \tilde \p_i  \tilde \p^j (E^a)_j
+  \, \epsilon^{a}_{~i} \tilde \p_j \p^k (E^j)_k 
- \frac{\bar c }2 \, \epsilon^{a}_{~i} 
  \epsilon_c{}^{j}  (E^c)_j . 
\end{split}
\end{equation}
To discuss the nature of linear excitations, 
let us write the EOM in the momentum space, 
\begin{equation}
  \omega^2 
  (E^a)_i 
  - 
  k^2 \, 
  \tilde n_i  (E^a)_j \tilde n^j
  - 
  k^2 \, \epsilon^{a}_{~i} \tilde n_j (E^j)_l n^l 
  - \frac{\bar c}2 \, \epsilon^{a}_{~i} 
  \epsilon_c{}^{j}  (E^c)_j 
  = 0 
\end{equation}
where $k \equiv \sqrt{k_i k^i}$, 
$\tilde k_i \equiv \epsilon_{ij}k^j$ 
is the transverse vector (to $k_i$), 
and 
$n_i \equiv k_i / k$, 
$\tilde n_i \equiv \tilde k_i / k$ 
are unit vectors in the longitudinal and transverse directions, respectively. 
In the long-wavelength limit, $k \to 0$, 
the antisymmetric part of 
the translational electric fields
$(E^a)_i$ satisfies 
\begin{equation}
\left( 
  \omega^2 - \bar c
\right) \epsilon_a{}^i (E^a)_i 
= 0. 
\end{equation}
Hence, 
as a result of the coupling to the rotational electric and magnetic fields, 
the antisymmetric part, 
$\epsilon_a{}^i (E^a)_i$, acquires a gap, 
and its gap is given by $\omega = \sqrt{\bar c}$. 
The other two modes are gapless. 
Because of the gap, the antisymmetric part 
$\epsilon_a{}^i (E^a)_i$ can be dropped at low energies \cite{Gromov:2019waa}, 
and the remaining electric field is symmetric. 
The symmetric tensor gauge theory~\cite{Pretko:2018qru}
is reproduced in this way.

To obtain the modes at finite $k$, 
let us project $(E^a)_i$ 
to the longitudinal and transverse directions 
for each index. 
We have the following projected components, 
\begin{equation}
TT \equiv \tilde n_a  (E^a)_i \tilde n^i, 
\quad 
LT \equiv  n_a  (E^a)_i \tilde n^i, 
\quad 
TL \equiv  \tilde n_a  (E^a)_i n^i. 
\end{equation}
As we stated earlier, $LL=0$ because of the Gauss law. 
Using the relation 
$\epsilon_{ij} = \tilde n_i n_j - n_i \tilde n_j$, 
we can write the antisymmetric part as 
$\epsilon_a{}^i (E^a)_i = TL - LT$. 
The EOMs are written in the matrix form as 
\begin{equation}
\begin{pmatrix}
  \omega^2 - k^2 - \frac{\bar c} 2 & k^2 + \frac{\bar c} 2 & 0  \\
  \frac{\bar c} 2 & \omega^2 - k^2 - \frac{\bar c} 2 & 0 \\
  0 & 0 & \omega^2 - k^2 
\end{pmatrix}
\begin{pmatrix}
  LT \\
  TL \\
  TT
\end{pmatrix}
= \bm 0  . 
\end{equation}
The transverse-transverse sector is decoupled 
and represents a gapless mode. 
There is a mixing between 
longitudinal-transverse and transverse-longitudinal sectors. 
As a result, there will be one gapless and one gapped mode 
from this sector.

Instead of electric fields, 
we can write the EOMs using the magnetic fields 
as follows: 
\begin{equation}
  \begin{split}
    (\p_0)^2 \mathsf{B}
   &=
    \tilde \p^i \p_0 \mathsf{E}_i 
    - 
    \frac{\bar c } 2  \epsilon_a{}^{i} \p_0 (E^a)_i   
   \\
   &= 
   \tilde \p^i \tilde \p_i \mathsf{B}
    - \frac{\bar c }2 \epsilon_{ai} 
    \left( 
       \tilde \p^i B^a +  \epsilon^{ai} \mathsf{B}
    \right)  \\
    &= \p^2 \mathsf{B}
    - 
    \bar c \,  \mathsf{B}
    + \frac{\bar c }2 \p_a B^a , 
  \end{split}
\end{equation}
\begin{equation}
  \begin{split}
    (\p_0)^2 (\p_a B^a) 
    &=
     \tilde \p^i \p_a \p_0 (E^a)_i  \\
    &=
     \tilde \p^i \p_a 
    [ \tilde \p_i B^a 
    + \epsilon^{a}_{~i} \mathsf{B} ]\\ 
    &= 
    \p^2 \p_a B^a -  \, \p^2 \mathsf{B} , 
  \end{split}
\end{equation}
\begin{equation}
  \begin{split}
    (\p_0)^2 (\tilde \p_a B^a)
    &= 
     \tilde \p^i \tilde \p_a \p_0(E^a)_i  \\
    &= 
     \tilde \p^i \tilde \p_a 
    [  \tilde \p_i B^a +  \epsilon^{a}_{~i} \mathsf{B} ] 
    \\
     &= 
     \p^2 (\tilde \p_a B^a) . 
  \end{split}
\end{equation}
Those equations can be written 
in the momentum space
in the following matrix form, 
\begin{equation}
  \begin{pmatrix}
   \omega^2 - k^2
   - \bar c 
   &  \bar c /2 &0 \\
  k^2 
  & 
  \omega^2 - k^2 
  &0 \\
  0 & 0 & 
  \omega^2 - k^2 
  \end{pmatrix}
  \begin{pmatrix}
    \mathsf{B} \\
    n_a B^a \\
    \tilde n_a B^a
  \end{pmatrix}
  = \bm 0 . 
\end{equation}
The longitudinal part of $B^a$ is mixed with 
the rotational magnetic field $\mathsf{B}$. 
The transverse part of $B^a$ is decoupled and stays gapless.

%
%
%
%
%
%
%
%
%
%
%

\end{CJK}

\bibliography{references}

\end{document}